\documentclass[aps,preprint]{revtex4}
\usepackage{epsfig}
\usepackage{wrapfig}
\begin{document}
\topmargin=0.2in
          
\title{Measurement of the Cosmic Ray Energy Spectrum and Composition from $10^{17}$ to $10^{18.3}$ eV Using a Hybrid Fluorescence Technique}
\author{T.Abu-Zayyad$^1$, K.Belov$^1$, D.J.Bird$^{5}$, 
 J.Boyer$^{4}$, Z.Cao$^1$, M.Catanese$^3$, G.F.Chen$^1$,
 R.W.Clay$^{5}$, C.E.Covault$^2$, 
 H.Y.Dai$ ^1$,
 B.R.Dawson$^5$, J.W.Elbert$^1$, B.E.Fick$^2$, L.F.Fortson$^{2a}$,
 J.W.Fowler$^2$, K.G.Gibbs$^2$, M.A.K.Glasmacher$^7$,
 K.D.Green$^2$, Y.Ho$^{10}$, A.Huang$^1$ , C.C.Jui$^1$, M.J.Kidd$^6$,
 D.B.Kieda$^1$, B.C.Knapp$^4$, S.Ko$^1$, C.G.Larsen$^1$,
 W.Lee$^{10}$, E.C.Loh$^1$, E.J.Mannel$^4$, J.Matthews$^9$,
 J.N.Matthews$^1$ , B.J.Newport$^2$, D.F.Nitz$^8$,
  R.A.Ong$^2$, K.M.Simpson$^5$, J.D.Smith$^1$,
 D.Sinclair$^6$, P.Sokolsky$^1$, P.Sommers$^1$, C.Song$^{10}$, 
 J.K.K.Tang$^1$, S.B.Thomas$^1$, J.C.van der Velde$^7$,
 L.R.Wiencke$^1$, C.R.Wilkinson$^5$, S.Yoshida$^1$ and
 X.Z.Zhang$^{10}$ }

\affiliation{$^1$ High Energy Astrophysics Institute,University of Utah, Salt
 Lake City UT 8 4112 USA\\$^2$ Enrico Fermi Institute, University of Chicago,
 Chicago IL 60637 USA\\ $^3$ Smithsonian Astrophys. Obs., Cambridge MA 02138
 USA\\ $^4$ Nevis Laboratory, Columbia University, Irvington NY 10533 USA\\
 $^5$ University of Adelaide, Adelaide S.A. 5005 Australia \\ $^6$ University
 of Illinois at Champaign-Urbana, Urbana IL 61801 USA\\ $^7$ University of
 Michigan, Ann Arbor MI 48109 USA\\ $^8$ Dept. of Physics, Michigan 
 Technological
 University, Houghton, MI 49931 USA\\ $^9$ Dept. of Physics and Astronomy,
 Louisiana State University, Baton Rouge LA 70803 and \\ Dept. of Physics,
 Southern University, Baton Rouge LA 70801 USA\\ $^{10}$ Dept. of Phys.,
 Columbia University, New York NY 10027 USA\\ $^a$ joint appt. with The Adler
 Planetarium and Astronomy Museum, Astronomy Dept., Chicago IL 60605 USA\\ }
 \date{\today}

 \begin{abstract} We study the spectrum and average mass composition of
 cosmic rays with primary energies between $10^{17}$eV and $10^{18}$eV using
 a hybrid detector consisting of the High Resolution Fly's Eye (HiRes)
 prototype and the MIA muon array.  Measurements have been made of the change
 in the depth of shower maximum as a function of energy. A complete Monte
 Carlo simulation of the detector response and comparisons with shower
 simulations leads to the conclusion that the cosmic ray intensity is changing from a heavier
 to a lighter composition in this energy range. The spectrum is consistent with
 earlier Fly's Eye measurements and supports the previously found  
steepening near $4\times 10^{17}$eV .
 \end{abstract} 
\maketitle 

 \section{ introduction} 
The source of cosmic rays with particle energies above $10^{14}$ eV is 
still unknown. Models of origin, acceleration, and 
propagation must be evaluated in light of the observed energy spectrum
and chemical composition of the cosmic rays.
Several experiments have attempted to determine the mean cosmic
ray composition through the ``knee'' region 
 of the spectrum, up to $3\times10^{16}$eV\cite{BLANCA}.  While
the results are not in complete agreement, there is some
consensus for a composition becoming heavier at energies above
the knee, a result consistent with
charge-dependent acceleration theories or rigidity-dependent
escape models.

In the region above the knee, the Fly's Eye experiment has reported a changing
composition from a heavy mix around $10^{17}$eV to a proton dominated flux
around $10^{19}$eV \cite{FEcomp}. This result makes this particular energy
region much more interesting than the expectation from a naive rigidity
model. This changing composition may imply that there may be multiple sources
of cosmic rays.   The  AGASA experiment shows broad
agreement with this trend if the data are interpreted  using the same
hadronic interaction model as used in the Fly's Eye analysis \cite{AGASA,DMS}.

The recently reported HiRes/MIA\cite{prl} hybrid observation on the cosmic ray
composition in a narrower energy region, $10^{17} \sim 10^{18}$ eV, shows a
general agreement with Fly's Eye experimental result. The new result gives a
somewhat more rapid change in the composition. The reliability of these
experimental results depends on how well we understand the development of all
components of extensive air showers (EAS) produced by cosmic rays, how well we
understand the response of our detector to the EAS, and how well we have done
in the EAS reconstruction. In this paper, we  address all
those issues in  detail.  We  start with a general description of the
techniques of cosmic ray composition measurement in the high energy region.

\subsection{Existing Techniques of Compostion Measurement above 10$^{17}$eV}

The Fly's Eye and AGASA experiments use different techniques to study
composition. The Fly's Eye experiment technique is based on the assumption
that the speed of the development of EAS depends on the mass of the primary
particle: a heavier nucleus induces earlier EAS development in the
atmosphere. The rapid break-up of a heavy nucleus at the early stage of the
cascade in the air leads to an effectively higher multiplicity than that
produced by a light nucleus or a proton at the same depth in the atmosphere. The
consequence is that the EAS is built up by a superposition of smaller
subshowers induced by nuclear fragments. In this case, since the subshowers
have lower energies, the EAS will have a shower maximum higher in the
atmosphere than in the case for a proton primary. The Fly's Eye experiment is
designed to measure the size, or total number of charged particles, 
of an EAS as a
function of atmospheric depth. It is thus an ideal detector to measure the
depth of maxima of showers directly.

In practice, the intrinsic fluctuations in the depth of shower maximum and the
detector resolution effects imply that one can not directly resolve the type
of primary nucleus on an event by event basis. 
 What one can do is to extract an average
EAS primary composition by comparing the data to a Monte Carlo simulation 
with a given primary composition. Since the shower development is
somewhat dependent on the choice of a hadronic model, this method leads 
to results which have some model dependence. 

The other method for studying composition depends on the assumption that the
muon content of EAS produced by a superposition of sub-showers ( as in the
case of heavy nuclei ) is larger than those with fewer sub-showers.  This is
due to the fact that the dissociation of a heavy nucleus produces a relatively
higher multiplicity in its interaction with atmospheric nuclei. The resultant
sharing of the primary energy between the nuclear fragments make the secondary
pions less energetic. As a consequence, those pions have a greater decay
probability into muons than those produced by a lighter nucleus in the early
stage of shower development. This leads to a difference in the muon content
between the EAS's induced by heavier and lighter nuclei, e.g. iron and
protons.  This difference shrinks with energy because the available path for
the decay of the high energy pions decreases as the shower develops deeper in
the atmosphere. According to the simulations the difference is still
resolvable in the energy region 10$^{17}$ $\sim$ 10$^{18}$ eV.

In principle , one
can obtain the information about the composition of the primary particle
by either measuring the total number of muons in EAS or a local
density of shower muons at a specific distance to the core of the shower.
However, as in the case of the previous method, the fluctuations are 
large compared with the separation between the different types of EAS, 
so that the
resolvability is not strong. The hadronic model dependence of the predicted
$\mu$-content from a particular composition is also significant for this
method.

\subsection{Advantages and Challenges for the HiRes/MIA Experiment}

The HiRes/MIA hybrid experiment is designed to combine the two methods
together using two independently developed co-sited experiments. The two
experiments share a trigger to simultaneously record both the longitudinal
development information and the EAS muon density.  This results in a
unique data set useful for the investigation of cosmic ray composition.  
Results on comparison of the flourescence and muon methods have previously been
published\cite{prl}. In this paper, we use the MIA array to improve the geometrical 
reconstruction and concentrate on the fluorescence technique
 for composition and spectrum studies.

The
hybrid timing information enhances the accuracy in the determination of the
geometry of the EAS. This accurate shower geometry plays a crucial role in the
subtraction of the Cerenkov light component of the EAS and in the corrections
for detector acceptance. These turn out to be the two key issues in the shower
longitudinal development profile determination. This profile provides directly
the size at shower maximum and its location while the integral of the profile
yields the shower energy.  The improvement in shower geometry determination
is the main advantage of a hybrid experiment.  The other advantage resulting
from the coincident measurement with the surface muon array is the existence
of a fully efficient triggering region in the central detector volume. This is
very helpful for the cosmic ray energy spectrum measurement.

One of the challenges for this experiment is that the two devices are
separated by only 3.3 km.  Because the MIA detector can not be triggered by
remote showers, most of the triggered events are located $\sim$ 4 km away from
HiRes detector.  This short shower-detector distance gives rise to difficulties
for reconstruction of those EAS's. The lateral distribution of shower
electrons is no longer a negligible effect.  The broadened source of light
increases the uncertainty in the shower geometry determination and in the
acceptance correction of the signals.  The limited effective trigger area of
MIA largely suppresses the aperture of HiRes detector. The energy
coverage of our data is  one decade of magnitude over 3 years of
observation, much smaller than that from the Fly's Eye experiment.  The
geometry of the triggered events is such that the fraction of Cerenkov light
is often large. This  implies the need for tight criteria for event
selection to reduce Cerenkov contamination, which in turn may cause a bias. We
must search for a balance between the tightness of the selection criteria and
the minimization of bias.  We describe our resolution of these problems 
 in the following sections.

\section{HiRes and CASA/MIA Experiments and Hybrid Observation}
In this experiment,  we use a hybrid detector consisting 
of the prototype High Resolution Fly's Eye (HiRes) air fluorescence 
detector and the Michigan Muon Array (MIA). 
The detectors are located in the western desert of Utah, USA
at $112^{\circ}\,\mbox{W}\,\mbox{longitude}$ and $40^{\circ}
\,\mbox{N}\,\mbox{latitude}$.  The HiRes detector is situated
atop Little Granite Mountain at a vertical atmospheric depth of
$860$ g/cm$^2$. It overlooks the CASA-MIA arrays some 3.4
km to the northeast. The surface arrays are some 150 m
below the fluorescence detector at an atmospheric depth of 
870 g/cm$^2$.

\subsection{ The HiRes Detector}
The HiRes prototype has been described in detail elsewhere
\cite{hires}. It views the night sky with an
array of 14 optical reflecting telescopes. They image the
EAS as it progresses through the detection volume from 3$^\circ$
 to 70$^\circ$ in elevation, 64$^\circ$ in azimuthal angle 
at the top and 32$^\circ$ at the bottom of the field of view.  Nitrogen
fluorescence light (in the 300--400\,nm band) is emitted at an atmospheric
depth $X$ in proportion to the number of charged particles in the
EAS at that depth, $S(X)$. The triggered tube directions and the 
time of arrival of light 
signals can be used to determine the shower-detector-plane 
and the tilt angle of the shower in this plane, denoted by $\psi$.
Part of the shower development profile
(at least 250 g/cm$^2$ long) can
be mapped by measuring the light flux arriving at the detector.
Assuming $S(X)$ to be the Gaisser-Hillas~\cite{gh} shower developement 
function and correcting for Cerenkov light contamination and
atmospheric scattering effects one can measure the
primary particle energy $E$, and the depth, $X_{max}$, at which the
shower reaches maximum size\cite{Baltrusaitis}.

\subsection{ The MIA Detector}
The MIA detector~\cite{casamia}, consisting of 16 patches 
formed with 64 scintillation 
counters each, covers about $370 \mbox{m}\times 370$ m with the active area over 
2500 m$^2$. The patches are buried 
about 3 m under the surface. The data acquisition system records
the identity and firing time of 
each counter participating in a given event. The EAS muon arrival times
are measured with a precision of 4 ns and all hits
occurring within 4 $\mu\mbox{s}$ of the system trigger are recorded. 
The average efficiency of MIA counters 
for detecting minimum ionizing particles was $93\%$ when they were 
buried, and the average threshold energy for vertical muons is about 
850 MeV. The MIA detector determines the muon density via the number and
pattern of hit counters 
observed in the shower \cite{MIAanalysis}.  

\subsection{ The Hybrid Trigger and Event Sample}
The HiRes detector collects data on clear moonless nights. A focal plane 
camera, consisting of 16$\times$16 photo-multiplier-tubes (PMT), is 
triggered if two of its 4$\times$4 ``sub-clusters'' contain at least 
3 fired tubes(two of them must be physically adjacent) in a 25 $\mu$s 
interval. Tubes trigger if the signal generates a voltage greater than 
a threshold, set at approximately 4 $\sigma$ above nightsky background noise.
 This yields a mirror trigger rate of about 30 to 120 
per minute. Once a trigger 
is formed, HiRes sends a Xenon light flash to MIA as a confirming 
trigger for a coincident event.

MIA has a 100\% duty cycle  and a  trigger rate of about 
1.5Hz formed by requiring at least 6 patches fired ( with at least 3 
hits found in each  patch). However, a coincident event is not
selected until either it is confirmed by a HiRes light flash 
communication signal if it is 
received within 50$\mu$s, or the event triggers CASA (a surface scintillation 
detector array for EAS electron observation\cite{casamia}) simultaneously 
and is coincident with a HiRes event within $\pm$3 ms according to the 
GPS clocks in each site. More details about the trigger formation 
and coincident event matching can be found in \cite{Brian}.

During the lifetime of this hybrid experiment between Aug. 23, 
1993 and May. 24, 1996 the total coincident
exposure time was 2878 hours corresponding to a duty cycle of
11.9\%. 4034 coincident events were recorded.
For events passing a set of coincidence assurance cuts
the shower trajectory, including arrival direction and core 
location for each event, was obtained in an
iterative procedure using the information from both HiRes and MIA
\cite{Brian}. 
2491 events survive this reconstruction procedure. 
 Further cuts are performed in 
order to maintain high resolutions in energy and shower maximum which 
are essential to the composition analysis. The criteria are  based on 
a thorough Monte Carlo simulation of the detectors as described below. 
After quality cuts, 891 events are employed in this composition 
study and energy spectrum measurement. 

\subsection{ Shower Geometry Determination}

Accurate knowledge of the shower geometry is  important in the
reconstruction of the shower profile in the atmosphere. The first step is to
use the pointing directions of triggered pixels in the HiRes detector to
determine the shower-detector-plane. This is a simple linear fit weighted by
the amount of light received by each tube.  The shower-detector-plane is
determined quite precisely because the HiRes detector records showers with an
average of 36 triggered pixels and the quality cut ensures that the lengths of
tracks is longer than 20$^\circ$. 
The typical error in the horizontal position of the plane is about 
0.1$^\circ$, while the error in the tilt angle of the plane is somewhat 
larger. The overall error in the shower-detector-plane normal direction 
is about 0.7$^\circ$.

Muon timing information from MIA plays a key role in  the
determination of geometric parameters in the 
shower-detector-plane, including the shower-detector distance and 
shower orientation, and  in the 
improvement of the shower-detector-plane determination. 
The initial trial shower arrival direction is determined 
 by fitting the muon arrival time with a flat shower front.
Projecting this direction onto the shower-detector-plane yields the  
tilt angle of the shower, $\psi$, defined as the angle between the 
shower axis and the horizon in the shower-detector-plane. 

\begin{figure}
\epsfig{file=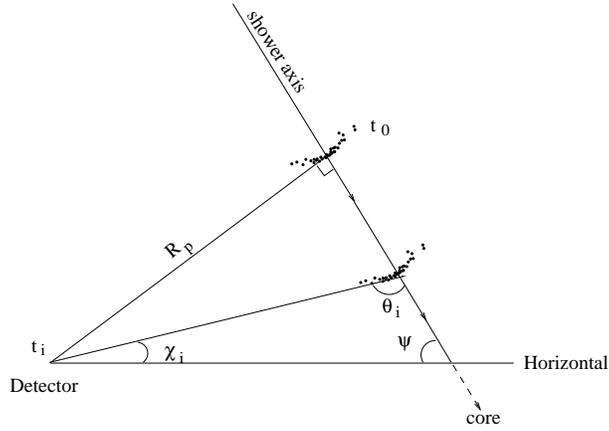,width = 8cm}
\caption{Illustration of shower geometric parameters in the  shower-detector-plane}
\label{sketch}
\end{figure}

Once $\psi$ is known, the light arrival time, t$_i$, on the $i$-th fired HiRes
tube is fit to the timing formula
\begin{eqnarray}
t_i = t_0 + \frac{R_p}{c} \mbox{ctan} \frac{\chi_i + \psi}{2},
\label{timing}
\end{eqnarray}
in which $\chi_i$ refers to the elevation angle of the  $i$-th 
fired tube. $t_0$ and $R_p$ are two parameters indicating the time 
 as the
shower front passes the detector and the perpendicular distance from 
the shower axis to the detector respectively, while $c$ refers to 
the speed of light( see the sketch in FIG.\,\ref{sketch}). 

\begin{figure}[t]
\epsfig{file=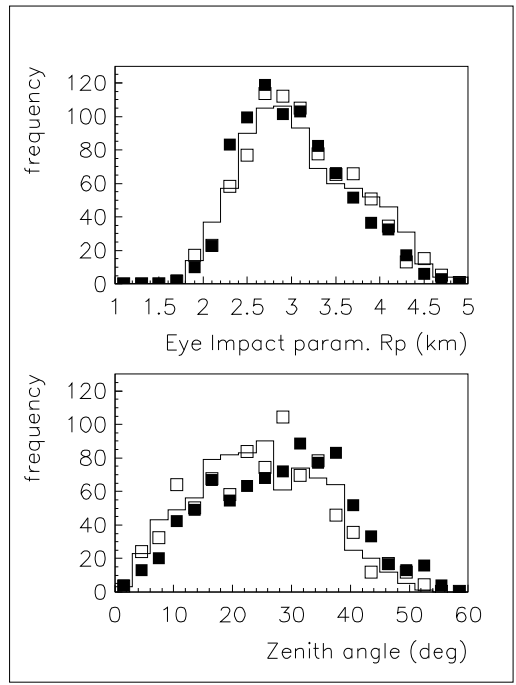} 
\caption{R$_p$ and zenith angle distributions. 
Filled squares are 
Monte Carlo predictions for proton, 
open squares for Fe (see text in next section for details).} 
\label{rp_dis} 
\end{figure}

\begin{figure}[t]
\epsfig{file=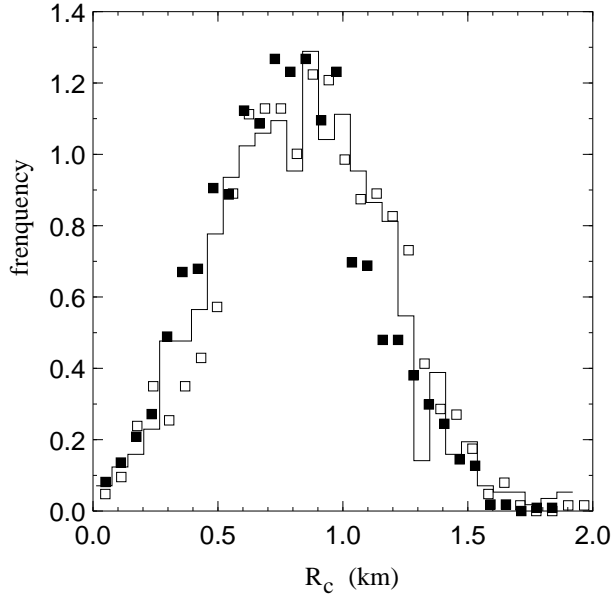, width=8cm}
\caption{Core location distribution with respect to the MIA center. 
Filled squares are 
Monte Carlo predictions for proton, 
open squares for Fe (see text in next section for details).} 
\label{core_dis}
\end{figure}

The core location can be derived by using the 
shower-detector-plane normal vector,
$\psi$ and $R_p$. The shower front shape can now be
more accurately  represented as a cone with
the delay parameter $\Delta=ar+br^2$, where $\Delta$ refers to the delay of
a conical muon front relative to the original flat front at a perpendicular
distance, $r$, to the shower axis.  This procedure is iterated after
additional corrections, including correction to the shower-detector-plane
direction. This iteration stops when the difference between the core
parameters is less than 10m.  The details of this iterative procedure can be
found in \cite{Brian}.  The distributions of those parameters are shown in
FIG.\,\ref{rp_dis} and FIG.\,\ref{core_dis}.  

\subsection{ Shower Longitudinal Development Reconstruction} 
The HiRes tube
signal, consisting mainly of the fluorescence light produced by charged
particles from the shower, can be used to reconstruct the 
shower development, i.e. to
calculate the shower size at the corresponding depth in the atmosphere. For this
purpose, the raw tube signal must be corrected for several effects.

If a triggered tube has a center that is not exactly in the shower-detector
plane, its signal requires correction for a number of effects. These 
 include the finite transverse width of the shower due to multiple
scattering of shower electrons, the 
finite size of the optical spot on the face of the
focal plane camera, the response function of the PMT cathode, gaps between the
pixels, and the effective light collecting area of mirrors.  All these effects
can be taken into account by performing a ``ray tracing'' procedure, namely we
trace the photons from the source direction, which can be wider than a line
source due to the lateral distribution of shower electrons, all the way down
to the face of the PMT via the spherical mirror surface. The response function
of the PMT cathode is folded into the ray tracing. The pixel signals are then
re-organized into a series of longitudinal ``bin signals'' with a 1$^\circ$
bin size along the shower axis. The signals are recalculated in units of
``number of photon-electrons per unit angle along the track per unit
collecting mirror area''. Ref. \cite{Brian} provides the details of this
correction and binning procedure for interested readers.

The fluorescence light yield (photons per meter ) of a single charged particle
varies slightly with the atmospheric pressure and temperature and has 
an energy dependence given by the dE/dX energy loss curve\cite{gene}. 
These effects are
taken into account in our shower reconstruction. The energy dependence of
the fluorescence light yield is folded in by taking an average over the shower
electrons at age $s$, defined as $3X/(X+2X_{max})$, with an energy
distribution of the shower electrons at age $s$ being extracted from the
simulation results using the CORSIKA package\cite{CORSIKA,chihwa}.  This energy
distribution is consistent with  measurements by Richardson
\cite{richardson} and a parameterization from Hillas \cite{Hillas}.

In addition to the corrections associated with those issues, the Cerenkov light
component of the bin signal must be subtracted as well, because only the pure
fluorescence light is proportional to the size of shower at a given depth.  
This gives rise to
some complications. First of all, the Cerenkov light component is composed of
the Cerenkov light produced by the shower electrons directly illuminating the
detector and the light scattered into the direction of the detector.  
Because the
Cerenkov light is very forward along the direction of the electrons, the
angular distribution of Cerenkov light in a shower is narrowly beamed and
falls exponentially with the angular distance from the shower axis.  The
average angular scale has been measured to be 
$4.0\pm0.3^\circ$\cite{BLANCA}.  Therefore, 
when the shower points towards the detector, the
estimate of the Cerenkov light becomes very sensitive to the geometry of
shower. In this case, the signal
is also dominated by  direct Cerenkov light.  We avoid
this sort of events in this analysis  by a cut on the minimal viewing
angle from the detector to the track, illustrated as the angle $\theta$ in
FIG.\,\ref{sketch}.

The scattered Cerenkov light, which we still have to deal with, includes
components due to the Rayleigh scattering from the atmospheric molecules and
Mie scattering from aerosols.  Rayleigh scattering is well understood in
terms of the distribution of scattering centers and their fluctuations, 
angular distribution of light, frequency response and the overall 
extinction length for ultraviolet
(UV) light.  This scattering is used to estimate the attenuation of Cerenkov
light along the shower, how much of this light is scattered into the PMT
direction, and the attenuation of the light during the propagation from the
source to the detector.

The scattering of UV light by aerosols is more uncertain. The distribution of
aerosols depends on weather conditions. Because of the variation  in the 
size distribution of aerosols, the scattering phase function can vary. 
There are several models for this function which lead to different
estimates of the amount of scattered light. We use the ``standard
desert aerosol  model'' with an exponential increase
of the aerosol extinction length with height above a mixing layer of height
$h_m$ below which the aerosol extinction length is constant. The change in
extinction length with height above this is governed by a ``scale height''
parameter $h_s$ while below $h_m$, the extinction length is given by the
``horizontal attenuation length'', $\lambda_a$. We monitor the aerosol
variation by the use of several Xenon flashers which shoot light pulses at
different angles into the atmosphere at different distances to the detector.
By detecting the scattered light from these flashers with the HiRes detector,
we monitor the variation of aerosols\cite{flasher} and partially constrain the
range of those parameters. This  will be discussed further in Sec. IV.

After subtracting  the Cerenkov light components, the bin signals, now
proportional to the size of the shower, are fit to a function describing the
longitudinal development of EAS suggested by Gaisser and Hillas\cite{gh}, (G-H
function): 
\begin{eqnarray}
 N(X) = N_{max} \left( \frac{X - X_0}{X_{max}-
X_0}\right)^{(X_{max}-X_0)/\lambda} e^{(X_{max}-X)/\lambda}, 
\label{ghfunc} 
\end{eqnarray} 
in which $N$ and $N_{max}$ refer to the size of the shower and its maximum, $X$
represents the atmospheric depth where the shower front passes the specific
angular bin, and $X_0$ and $X_{max}$ the depths where the shower starts and
reaches its maximum, respectively. The depth $X$ is calculated based on the
shower geometric parameters and the pointing directions of each effective
angular bin. The choice of this EAS
longitudinal development curve has recently been experimentally 
confirmed \cite{profile} as
accurate based on the same data set.  The G-H function  is one of the
best parameterizations of longitudinal development according to this study.

\begin{figure}[t] 
\epsfig{file=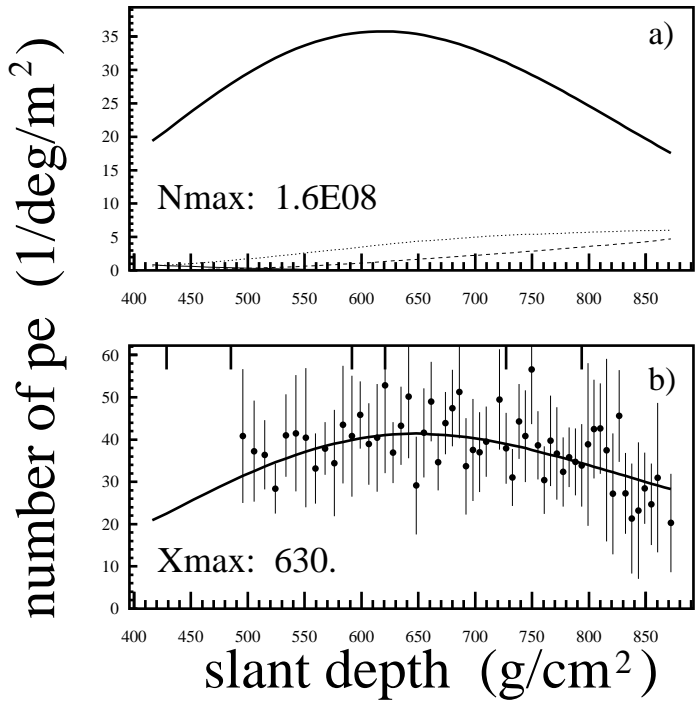, width=10cm} 
\caption{A
typical event. See the text for details.}  
\label{example} 
\end{figure}

An example of shower reconstruction is shown in FIG.\,\ref{example}.  In a),
the four components of light contributing to the best fit results are plotted:
fluorescence light (thick solid line), direct Cerenkov light (thin solid
line), Cerenkov light from Rayleigh scattering(dotted line) and Cerenkov
light through aerosol scattering (dashed line). In b), we show the fit of the
sum of all the components to the bin signals(dots). This shower reaches its
maximum size of 1.6$\times$10$^8$ at a depth of 630 g/cm$^2$.  
Note that the parameter $X_0$ is
the point at which size $N=0$ according to (\ref{ghfunc}). We  are not
sensitive to light from early shower development and we fix $X_0$ at -20 g/cm$^2$
in our fitting procedure. The justification for this will be given below.

\subsection{ Shower Energy Determination}
Once we know the longitudinal development profile of an EAS, we can 
integrate over its depth to calculate the total path length
of all shower charged particles and calculate the total
deposited energy $E_{e.m.}$ by the charged particles in  
this shower, i.e.,
\begin{eqnarray}
E_{e.m.} = \frac{E_c}{L_0}\int N(X)dX,
\label{Eem}
\end{eqnarray}
where the critical energy and the radiation length of electrons in  air 
are $E_c$ and $L_0$, respectively. A recent study\cite{chihwa}
 based on the Monte Carlo simulation package CORSIKA verifies
this formula and re-evaluates the constant $E_c/L_0$ as 2.19. 
Since some of the primary energy is carried away by neutrinos and muons 
penetrating  the ground, a correction for this effect must be applied.
In that 
study, the authors establish a new empirical formula 
for the converting $E_{e.m.}$ into total energy of the shower, $E_0$. It 
reads 
\begin{eqnarray}
E_0 = \frac{E_{e.m.} }{A - B E_{e.m.}^{\kappa} }
\label{E_0}
\end{eqnarray}
where the parameter $A=0.959\pm 0.003$, $B=0.082\pm 0.003$ and 
$\kappa=-0.150\pm 0.006$. These parameters are determined by taking an 
average between proton and iron initiated showers, since it is impossible 
to know the primary particle mass in advance of 
the reconstruction. This causes a
systematic uncertainty in $E_0$ of less than 10\%. 

\begin{figure}[t]
\epsfig{file=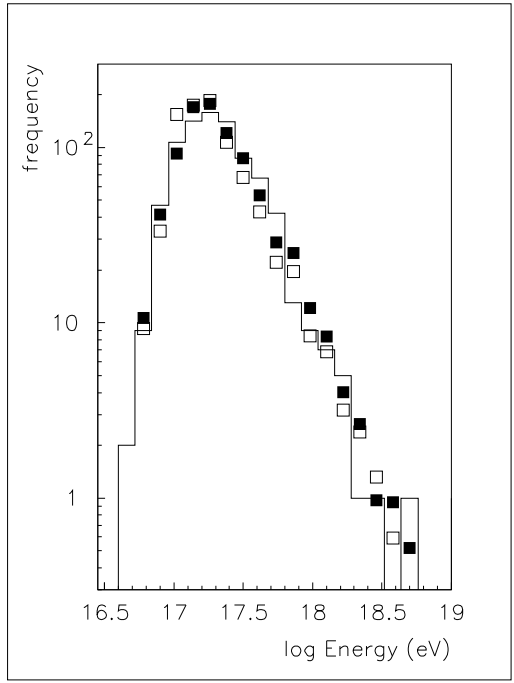}
\caption{Energy distribution. The histogram represents the data 
and the dots represent simulations, see text in Section IV for details.}
\label{e_dis}
\end{figure}

The energy distribution of our data set is displayed in FIG.\,\ref{e_dis}.
The vertical axis represents the
number of events within a bin of $log_{10}$E. The figure shows  that
the threshold of our detector is about 5$\times$10$^{16}$ eV.
The hybrid detector approaches a fully efficient operation
 above 4$\times$10$^{17}$ eV.  

\section{Monte Carlo Study of Detector Resolutions}
In order to test
this complex EAS reconstruction procedure and evaluate the resolution in
shower geometry, shower depth of maximum and energy based 
on this reconstruction scheme, we
have developed a Monte Carlo code to simulate the EAS shower and 
detector. We have made this
as realistic as possible both for shower development in the atmosphere and for
the response of our detector to the shower. In this section, we will address
how this event generator is driven with a full Monte Carlo simulation of
EAS, how the production and propagation of light through the atmosphere
is treated, how the acceptance and response of the detector to the light is
simulated, and what the final resolution functions and their relationship with
the event quality cuts are.

\subsection{ Shower Generation: CORSIKA Package and Hadronic Interaction
Models} The driver of the shower simulation is a series of parameterizations
of the results from a full EAS simulation using the CORSIKA
package\cite{CORSIKA}.  This is one of the most modern and complete simulation
codes for EAS development. It traces shower particles from very high energy at
top of the atmosphere down to the threshold energy of 100 keV. A ``thinning''
technique is used in the simulation to reduce the size of the
calculation. Only one secondary particle is traced if the interaction energy
falls below the thinning threshold, e.g. 10$^{-5}$ of the shower total
energy. A weight is assigned to this traced particle to represent those not
being traced.  Depending on the degree of realism in the fluctuations that is
required, the user can set an appropriate thinning threshold if the CPU time
limit allows.  Another advantage of the CORSIKA code is that the user can
switch between several optional hadronic interaction models.  The
authors have made efforts to test the program for  primary particle
energies up to 10$^{16}$ eV, but do not claim reliability for energies
higher than 10$^{17}$ eV. However, it is one of the best EAS models currently
available.  Low energy shower particles, down to the tens
of keV level, are treated  carefully by employing the well known
EGS package \cite{EGS} etc.

In our simulation with CORSIKA Ver. 5.624, the thinning threshold 
is set to be 10$^{-5}$ of the shower energy and the QGSJET\cite{QGSJET} 
and SIBYLL\cite{SIBYLL} hadronic interaction models are selected. The 
number of EAS electrons as a function of depth and EAS muon information,
including arrival direction, time and energy for every muon above 
870/cos$\theta$ MeV at 870 g/cm$^2$, are recorded. We simulated 500 
events for each of the 5$\times$4 grid points in energy from 
3$\times$10$^{16}$ eV to 5$\times$10$^{18}$ eV and zenith angle from 
0$^\circ$ to 60$^\circ$. The same number of events were generated for
proton and iron induced showers and under different hadronic interaction 
model assumptions.

\subsection{ Shower Longitudinal Development Profile Parameterization} 
\begin{figure}[t] 
\epsfig{file=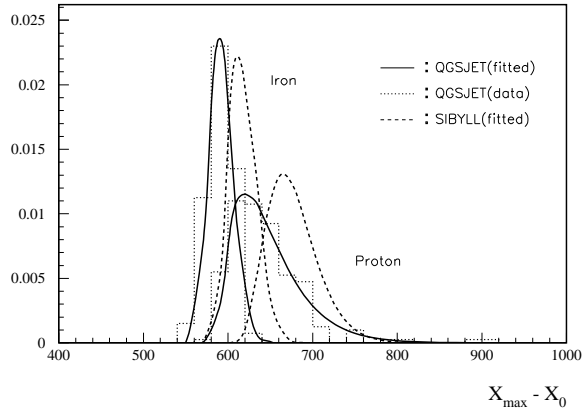, width=9cm} 
\caption{The fluctuation of $X_{max}$ and its parameterization at 
$2\times 10^{17}$ eV. }
\label{xm_mc_dis} \end{figure}

Based on this large simulated event data set, we parameterized all the 
distributions
of EAS parameters such as the first interaction depth $X_1$, shower decay
constant $\lambda$, shower maximum $N_{max}$, its position in depth
$X_{max}$, and the correlations between them. As an example, the
$X_{max}$-distribution is shown in FIG.\,\ref{xm_mc_dis}. It is clear that the
proton induced showers possess larger fluctuation in $X_{max}$ than iron
initiated ones.  Though they overlap each other, the means for each
distribution are about 100 gm/cm$^2$ apart, 
which is resolvable if sufficient
statistics are available.  The model dependence appears to be significant, but
is smaller than the proton-iron separation.  The comparison between the
histogram for simulated data and curves for parametrized results shows that
the parameterization faithfully represents the fluctuation in $X_{max}$. A
similar situation is found for the other parameters. Among the parameters,
we find that $\lambda$ and $N_{max}$ are correlated and we put this
correlation into our generator.  

We find that the parameter $X_0$ is quite insensitive to the type of primary
particle and energy if we use the G-H function (\ref{ghfunc}) to fit the
longitudinal development of simulated showers.  The fitting quality remains
quite good by fixing it at a value of -20 g/cm$^2$. The other parameter
$\lambda$ is found to have a slow variation with energy and mass of the
primary particle, with a central value of 70g/cm$^2$.  We fix both at the
values suggested here in our reconstruction procedure for real events. One of
the benefits of fixing those relatively insensitive parameters is to reduces
the chances of the parameter search being trapped at a local minimum of
$\chi^2$.

Once the shower parameters are determined as a function of 
energy, the number of electrons can be calculated by using the G-H function 
(\ref{ghfunc}) at depth $X$. The electrons are distributed laterally 
according to the NKG function  at corresponding age $s$ of the shower.
The fluorescence and Cerenkov 
light and the corresponding signal appearing at the HiRes detector can 
then be generated. 

\subsection{ Muon Lateral Density and Arrival Time Distribution.}  
\begin{figure}[t] 
\epsfig{file=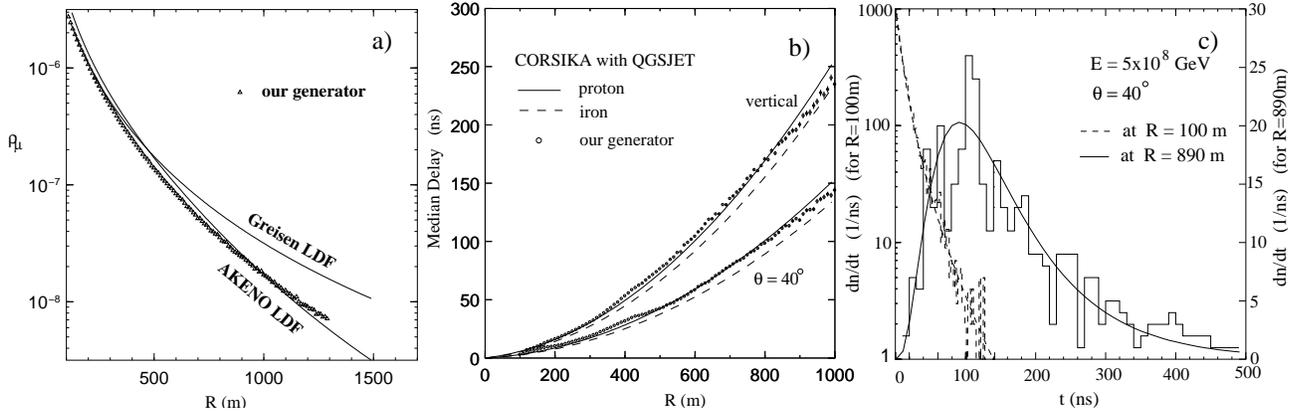, width=17cm}
\caption{Lateral distribution of $\mu$-density, shower $\mu$-front and time
structure of $\mu$-disk at 870 g/cm$^2$} 
\label{muon} 
\end{figure}

The
simulation of muons in an EAS is much more complex. The
dependence on both zenith angle and energy is important 
since the observation is done at a
fixed altitude. In order to simulate the MIA trigger correctly, we generate
the muon density, $\rho_\mu(R)$ at a distance $R$ to the core according to the
muon lateral distribution.  $\rho_\mu(R)$ and its fluctuation behavior are
parameterized based on simulations. The muon density generated based on our
parameterization is plotted in the FIG.\ref{muon} a). As a comparison, the
AGASA muon lateral density function(LDF)\cite{AGASA} and the Griesen LDF are
plotted in the same figure. Our simulation agrees with the AGASA LDF well 
except in
the small core distance area in which our simulation is closer to the Griesen
function. 

The arrival time of the EAS muon is essential in the simulation for MIA
triggering. We parameterize the distribution of arrival time for each muon in
the shower disk within different annular rings at a distance 
$R$ from the core, and
at all zenith angles and energies in the grid mentioned above.  
The shape of the arrival time distribution changes quite rapidly 
with $R$ as can be seen in
FIG.\,\ref{muon}\,c). Note the vertical scales are different for the
different cases.  We use a single function 
\begin{eqnarray} 
\frac{dN}{dt}
\propto t^\alpha \mbox{exp}\left\{\frac{-t^\beta}{\tau}\right\}, 
\label{dt}
\end{eqnarray} 
to describe all these distributions.  The parameters
$\alpha$, $\beta$ and $\tau$ are tabulated as functions of $R$, energy and
zenith angle. The parameters are generated with due regard to correlations if
they exist and the muon arrival time are generated individually
depending on how many muons are generated at $R$ and for a 
specific direction.  The
median time for muons generated in an annular ring at $R$ reveals the
curvature of the shower front.  In figure b), we plot two examples at zenith
angle $\theta=0^\circ$ and $40^\circ$. The small differences between proton
and iron induced showers are also shown in the figure. The lines represent the
results directly from CORSIKA simulation and the dots are from our proton
shower generator.  

\subsection{ Detector Response and Resolution} 
Once the shower is generated
by the CORSIKA based driver, fluorescence and Cerenkov light contributions are
calculated for each  0.04$^\circ$
angular bin along the axis of the shower.  Cerenkov light generated
in the previous bins is accumulated  taking account of the attenuation
between the bins due to the scattering of the light on atmospheric molecules
and aerosols. The propagation and attenuation of light through the air and
acceptance of the light by the detector are simulated in detail. The effective
area and reflectivity of the mirror, filter transmission on the face of
the focal plane camera, quantum efficiency of the PMT cathode, 
the gain of PMT and all
electronics triggering and charge integration are realistically simulated. The
variation of all these effects with wavelength is considered by tracking the
light in 16 different UV wavelength bands covering from 300 to 400
nm. Finally, the simulated signal in each pixel is built up by summing all
the angular bins involved and the 16 wavelength bands. The night 
sky background light
is also added to the signal according to results of an on site
measurement\cite{noise}.  The trigger time slewing (i.e. late triggering for
small pulses) is also taken into account for the HiRes sample-and-hold
electronics.

The MIA signal is generated by sampling the number of muons for each counter
at the perpendicular distance $R$ to the shower axis, then run over all 
muons in 
each counter to generate the arrival time for each of them. The pulse 
from a given counter
is built up by passing all the muon signals 
through the electronics sequentially. 
Dead counters, counter efficiencies, trigger formation, time windows 
for accepting counter hits and noise muons are taken into account in the 
simulation. Both the 
HiRes and MIA signals are written in the same format as for the real data.

\begin{figure}[t] 
\epsfig{file=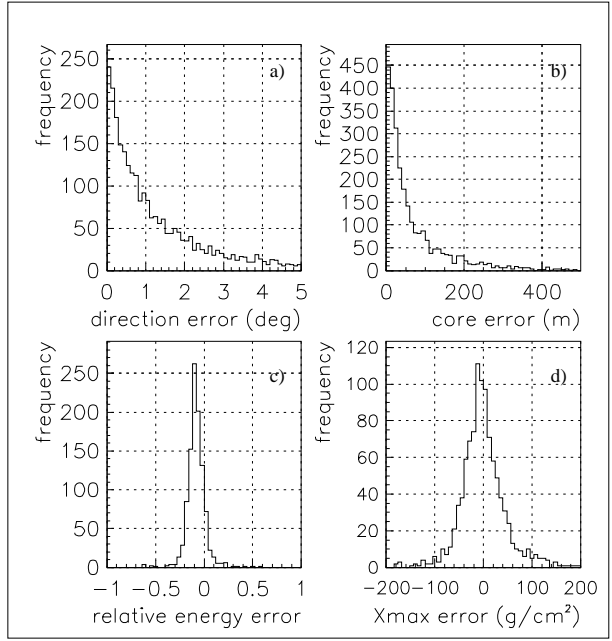, width=8cm} 
\caption{Resolution
functions of shower arrival direction a), core location b), shower total
energy c) and depth of shower maximum d).  The differences between the
generated values of those variables in MC and those obtained from the
reconstruction of the detector output are plotted here. 
Energy error is in $(E-E_{in})/E_{in}$ in c), where $E_{in}$ is the 
input from the simulation.}  
\label{resolutions}
\end{figure}

8000 proton and 4000 iron induced showers are generated with a spectrum of
trial energies from $5\times10^{16}$ to $5\times10^{18}$eV. The differential
spectral index is set as -3.0. All the simulated events are passed through the
same reconstruction procedure as real data and all geometrical and shower
development parameters are determined. Events must pass the ``quality cuts" 
defined below.   We compare
the overall distributions from the simulated events with data. This is plotted
in FIG.\,\ref{rp_dis},\ref{core_dis},\ref{e_dis}. The solid(open) squares
represent the showers induced by protons(iron nuclei) in those figures.  The
consistency between the data and simulation builds up our confidence in the
simulation, and hence in the resolution functions we now present. 
Since we know the input parameters for
every event, we can study the detector response and the corresponding
resolution function on an event by event basis. In
FIG.\,\ref{resolutions}, the resolution functions in shower arrival direction,
core location, energy and $X_{max}$ are plotted for iron induced showers. In
a), we plot the distribution of the ``opening angle'' between the ``real''
shower axis vector given by the input shower and the ``reconstructed'' shower
axis vector which is determined by using of the timing information from both
HiRes and MIA. Similarly in b), we plot the distribution of the distance
between the input shower core and the reconstructed core position. Those two
geometrical resolution functions peak at zero but the long tails imply that
some events are measured  poorly. 

We now move to a more quantitative discussion of parts c) and d) in the
figure, the resolution functions in energy of primary particle and shower
maximum depth, respectively. These are the most important results for the 
energy spectrum 
measurement and the composition study described in this paper.

\subsection{ Hybrid ``Good'' Event Criteria versus Resolution}

The better the geometrical parameters of a shower are determined, the better
the shower development profile can be extracted.  Of all geometrical
parameters, the shower-detector-plane is the most crucial and depends strongly
on how many tubes are triggered and how long the track formed by those tubes
is. The number of muons detected by MIA is the other contributor to precise
time fitting.  In order to locate the shower maximum, it and a
good fraction of the rest of the profile must bee seen by the
detector. Moreover, as mentioned before, we must avoid those events 
which are dominated by
Cerenkov light. Poorly fitted events should also be
rejected. The set of quality cuts listed in
Table.\,\ref{cut_tab} addresses these issues.  
\begin{table} 
\begin{tabular}{|c|c|} 
\hline Variables
& cuts \\ 
\hline
\hline 
Track Angular Length & $>20^\circ$ \\ 
\hline 
$R_{p_{MIA}}$ & $<$2 km \\
\hline 
$X_m$ & $X_l < X_m < X_h$\\ 
\hline 
Spanning & $X_h - X_l > 250$
(g/cm$^2$) \\ 
\hline 
$\theta_h$ & $>10^\circ$\\ 
\hline 
$\Delta X_m$ & $<50$(g/cm$^2$)\\ 
\hline 
$\chi^2$ per DOF & $<$10 \\ 
\hline 
\end{tabular} 
\caption{Event
criteria. $X_h$($X_l$) refers to the depth of the highest(lowest) section of
the shower track seen by the detector. $\Delta X_m$ is the estimated error
in $X_{max}$.}  
\label{cut_tab} 
\end{table} 
The first four criteria in this table are self explanatory 
while the fifth throws out
events which come towards the detector and are dominated by Cerenkov
light. This cut reduces the Cerenkov light 
contribution to less than 75\%  of the total
amount of light in each event.  The average Cerenkov light is about 25\%.
The last two cuts control the fitting quality.

After these tight cuts, the means and widths of the resolution functions, as
shown in FIG.\,\ref{resolutions}, are summarized in Table.\,\ref{res_tab}.  
The resolution is significantly improved compared with the Fly's Eye
experiment. In that experiment 
the energy resolution was 33\% (monocular) and 24\%
(stereo) below 2$\times$10$^{18}$eV\cite{apj} and the $X_{max}$ resolution was
 50 g/cm$^2$ averaged over a broader energy range up to
10$^{19}$eV\cite{FEcomp}.  Most importantly, there is
no bias observed in the present experiment with these cuts. The resolution
functions show negligible systematic shifts except in energy. Those
shifts go in opposite direction for proton and iron induced showers. This is
discussed further in Sec. II F.

\begin{table}
\begin{tabular}{|l|c|c|c|c|}
\hline
QGSJET & \multicolumn{2}{c|}{proton} & \multicolumn{2}{c|}{iron} \\
\hline
& \hspace{2mm}$\sigma$\hspace{2mm} & mean & \hspace{2mm}$\sigma$\hspace{2mm} & 
mean \\
\hline
E (\%) & 16 & 8 & 10 & -13 \\
\hline
$X_{\rm max}$ (g/cm$^2$) & 44 & 7 & 44 & -2 \\
\hline
$X_{core}$ (m) & 42 & -2 & 40 & -1 \\
\hline
$Y_{core}$ (m) & 57 & -2 & 55 & 2 \\
\hline
space angle & \multicolumn{2}{c|}{$0.88^\circ$} & \multicolumn{2}{c|}{$0.83^\circ$} \\
\hline
\end{tabular}
\caption{Resolution figures for a $E^{-3}$ differential spectrum seen
by HiRes and MIA.  Quality cuts have been applied.
Space angle errors are median values.}
\label{res_tab}
\end{table}

The other important issue is the energy dependence of
those resolution functions. Because of the constraint from MIA, all the well
reconstructed showers are at a similar distance from the HiRes detector. The
energy and $X_{max}$ variables, which are mainly determined by HiRes, thus have a
resolution function which varies slowly with energy. They appear slightly
worse at 10$^{17}$eV 
 due to closeness to the  detector threshold.  The width of energy
resolution shown as FIG.\,\ref{resolutions} c) changes from 11\% at
10$^{17} eV$ to 6\% at 10$^{18}$ eV for iron induced showers.  The $X_{max}$
resolution changes from 48 g/cm$^2$ to 41 g/cm$^2$ over the same range.

In summary, we have established the validity of our full Monte Carlo code for
the HiRes/MIA detector and for the reconstruction procedure. We evaluate the
resolutions for all interesting variables and optimize the resolution by
selecting ``good'' events with a set of tight cuts which do not cause
bias. Under these criteria, 891 real events remain. They form
the data base of our measurement of energy spectrum
 and investigation of composition of
cosmic rays in the energy range covered by this data set.

\section{ Physics Results}
We measure the cosmic ray intensity as a function of energy and study the 
cosmic ray composition in the energy range from $10^{17}$ to $3\times 10^{18}$
eV using this experimental data set and the detailed  Monte Carlo study.
All the results are summarized in this section.

\subsection{ Aperture Estimation } 
As shown in FIG.\,\ref{e_dis}, the shower energy distribution peaks 
at $3\times 10^{17}$ eV,  which points to a  fully  efficient observation above
$4\times 10^{17}$ eV. The distribution falls off rapidly 
below $3\times 10^{17}$ eV
due to trigger inefficiency. In order to measure the cosmic ray  
energy spectrum, a correction for the detector aperture is necessary.  

There are two ways to estimate the aperture of the hybrid detector. One way
is to use the full simulation code, calculate the observation efficiency 
with the same reconstruction, and event selection criteria at several fixed 
energies in this energy region. The other way is to use the 
measured core location  and arrival direction distributions of the 
observed events directly . 

\begin{figure}[t]
\epsfig{file=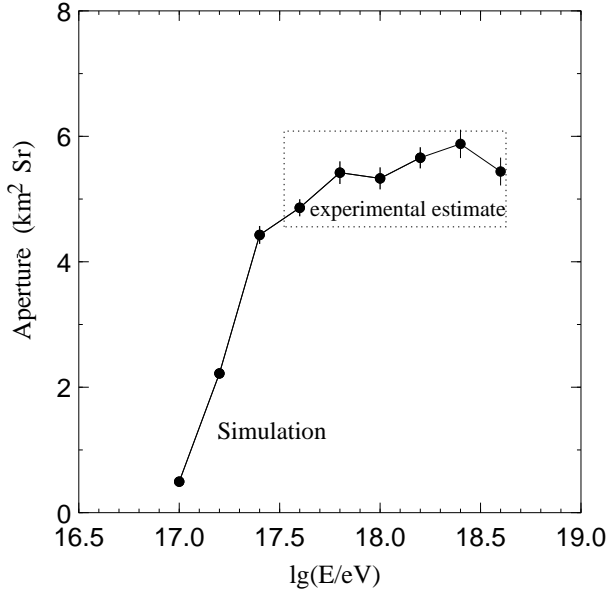,width=8cm}  
\caption{The detector aperture as a function of primary energy. Dots 
connected by solid line represent the simulated result. The area marked
by the doted line gives a range of the experimentally estimated aperture
with its uncertainty. } 
\label{aperture}
\end{figure}

Above a certain energy, the hybrid detection scheme becomes fully efficient
and the aperture should exhibit a plateau for events near the center of the
detection volume and within a given solid angle.  
This indicates that the detector is fully efficient in this kernel
of the detection volume. The hieight of the plateau provides a normaliztion 
reference for the distribution of detector efficiency as a function of 
solid angle and location within the detection volume. 
 The detector aperture can be estimated by integrating this normalized
efficiency distribution over the whole area and $2\pi$ solid angle.
In our case, we find that this kernel corresponds to an area with 0.8 $km$
radius centered at  MIA and a cone with 30$^\circ$ zenith angle. For a 
distribution of 182 events with energy higher than $10^{17.6}$ eV, 
the integral yields an aperture of about 5.2 $ km^2\cdot Sr$. 
By using this method, one
can avoid the modeling dependence inherent in the MC simulation. However, poorer
statistics will result a large uncertainty in the aperture.
FIG.\,\ref{aperture} shows the results. The area surrounded by the dotted line
indicates this experimentally estimated aperture with an uncertainty of $0.8
km^2\cdot Sr$ dominated by statistic error. 

The Monte Carlo method provides a more precise estimate. Every point in the
figure represents about 3,000 events. Above $10^{17.6}$ eV, the calculation
shows that the detection efficiency is saturated and this aperture is
consistent with the experimental estimate.  The feature of a flat aperture as
a function of energy, provided by the MIA detector, is very
useful for the cosmic ray intensity measurement.  The price for this feature is
that the aperture is rather small. The MC method  provides a
calculation of the aperture near detector threshold with good precision. 
 The fluorescence detector has a sharp threshold around
$10^{17}$ eV. Since the efficiency drops to lower than 10\%, the events 
 below this energy are not included.

\subsection{Energy Spectrum} 
\begin{figure}
\epsfig{file=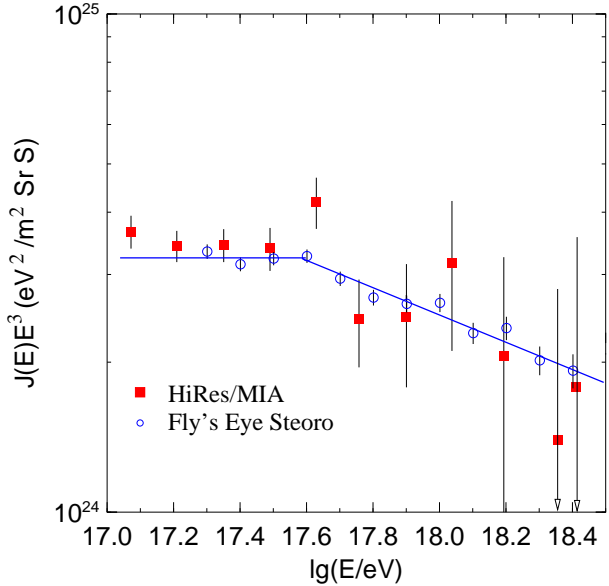,width=8cm} 
\caption{The differential energy spectrum of cosmic rays in the vicinity of 
$3\times 10^{17}$ eV. $E^3$ is multiplied to the intensity. 
The result from this experiment (squares) is consistent
with the Fly's Eye experiment (dots). The lines represent the fit according to 
Fly's Eye data\cite{FEcomp}.} 
\label{differential} 
\end{figure}

FIG.\,\ref{differential} shows the cosmic ray
energy spectrum from $10^{17}$ eV to $2.5\times 10^{18}$ eV. The
total exposure is about 1.45$\times 10^{13}$ $m^2\cdot Sr\cdot sec$. In order to see the
detailed structure of the energy spectrum, the intensity is multiplied by
$E^3$. Due to the small exposure, statistics are poor above $3\times 10^{17}$
eV. Nevertheless, the data supports an overall power law spectrum with 
an index about -3.10
and a intensity of $10^{-29.45} eV^2\cdot m^{-2}\cdot Sr^{-1}\cdot s^{-1}$ at
$10^{18}$ eV. The data of this spectrum is listed in Table III. 
\begin{table}
\begin{tabular}{|c|c|c|}
\hline
\hline
log$_{10}$ (E/eV) & J(E) & $\Delta$ J \\
\hline
   & ($10^{-28}\cdot eV^{-1}\cdot m^{-2}\cdot Sr^{-1}\cdot s^{-1}$) & 
          ($10^{-28}\cdot eV^{-1}\cdot m^{-2}\cdot Sr^{-1}\cdot s^{-1}$) \\
\hline
17.07& 22.3& 1.7\\
\hline
17.21& 8.04& 0.58\\
\hline
17.35& 3.05& 0.23\\
\hline
17.49& 1.15& 0.11\\
\hline
17.63& 0.542& 0.064\\
\hline
17.76& 0.130& 0.026\\
\hline
17.90& 0.049& 0.014\\
\hline
18.04& 0.0244& 0.0081\\
\hline
18.20& 0.0054& 0.0031\\
\hline
18.36& 0.0012& 0.0012\\
\hline
18.41& 0.00105& 0.00105\\
\hline
\hline
\end{tabular}
\caption{The cosmic ray energy spectrum from $10^{17}$ eV to $3\times 10^{18}$
eV.}
\label{listtab}
\end{table}

As a comparison, the stereo 
Fly's Eye \cite{FEcomp} measured energy
spectrum is plotted in the same figure. The new measurement is consistent with
this Fly's Eye result. The difference between the two measured intensities
 is less
then 8\% below $3\times 10^{17}$ eV.  

The new measurement marginally confirms the
energy spectrum break occurring around $3\times 10^{17}$ eV in the old Fly's
Eye stereo data. This break is not seen in the Fly's Eye monocular data set
because of poorer energy resolution. Since the HiRes/MIA experiment has 
even better energy resolution,  it should see this break if it
exists.

\begin{figure}
\epsfig{file=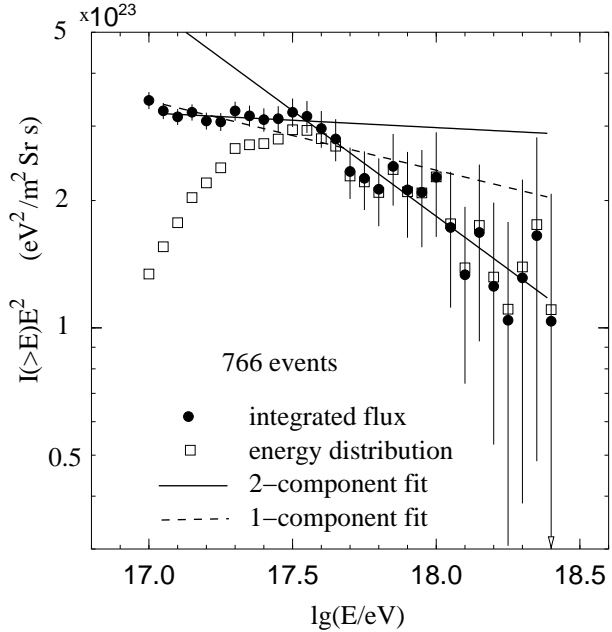,width=8cm}
\caption{The integrated energy spectrum of cosmic rays measured with
this experiment (dots). $E^3$ is multiplied to the intensity.
Open squares show the energy distribution before 
the aperture correction. Solid lines represent a fit of power law with two 
indices and dashed line shows an overall fit.}
\label{integral}
\end{figure}

In order to avoid possible binning bias inherent with such low statistics, we
plot the integrated energy spectrum in FIG.\,\ref{integral}. The break in the
spectrum is clear. By using a least $\chi^2$ fit, we find that the break
occurs around $10^{17.5}$ eV. The spectrum parameters are summarized in the
table.\,\ref{spectab}.  The raw energy distribution is plotted in the same
figure to demonstrate the effect of the detector aperture correction.  Above
$10^{17.6}$ eV where the aperture is flat, the absolute value of the spectrum
index is too large to be consistent with the lower energy measurements
\cite{BLANCA}. The energy spectrum below $10^{17.3}$ eV 
requires a large correction but
it is significantly different from the power law fit to the correction-free
data above $10^{17.6}$ eV and in good agreement with lower energy experiments.

\begin{table}
\begin{tabular}{|l|c|c|c|}
\hline
\hline
Energy (eV) & Index & lg(J(10$^{18}$ eV)) \\
\hline
$10^{17.0} \sim 10^{17.5}$ & $-3.07\pm 0.11$ & -29.45 \\
\hline
$10^{17.5} \sim 10^{18.4}$ & $-3.52\pm 0.19$ & -29.55 \\
\hline
\hline
\end{tabular}
\caption{The parameters of cosmic ray energy spectrum. 
Both components are listed
with their energy range. The normalization of the cosmic ray intensity,
J,
is provided at $10^{18}$ eV for both cases, in 
$m^{-2}\cdot Sr^{-1}\cdot s^{-1}\cdot eV^{-1}$.}
\label{spectab}
\end{table}

\subsection{Resolvability of Composition: $X_{max}$ and $E.R.$} 
As mentioned in the introduction, the
distribution of $X_{max}$ can be used 
with the help of Monte Carlo simulations to extract an average composition of
primary cosmic rays. The results
depend on the hadronic interaction model, 
the EAS simulation model and resolution
of the detector in energy and $X_{max}$.  In this paper, we compare data to
the predictions of two hadronic models: QGSJET and SIBYLL. Other, lower
multiplicity models have been shown to be inconsistent with any normal
composition of cosmic rays.

According to the simulation, the average $X_{max}$ for showers induced by
protons is separated from the average for iron by about 100 g/cm$^2$ and this
separation is almost independent of energy in the range from 10$^{17}$eV to
10$^{18}$eV. Since the resolution of the detector is 44g/cm$^2$, we can
tell if the data is closer to one than the other.  
The absolute position of $\overline{X_{max}}$ for a given composition
assumption is model dependent (about 25 g/cm$^2$ shift). At any 
given energy, a measurement of  $\overline{X_{max}}$ implies a 
particular composition which is thus 
systematically uncertain. On the other hand, an apparent
departure of the data points from either of the predictions based on proton or
iron showers as a function of energy will reveal information on the change 
in the composition of cosmic rays. The rate of this change can be much more 
reliably determined than the absolute composition itself. 

The variation of the separation between the measured and 
simulated pure composition $X_{max}$ can be quantitatively 
evaluated using the so
called ``elongation rate''.  Based on models, it is found that the average
$X_{max}$ increases with energy logarithmically over the energy range of
interest.  The ``elongation rate'' is symbolized by $\alpha$ in this paper 
and defined by 
\begin{eqnarray}
\overline{X_{max}}  \propto \alpha \mbox{log}E.
\label{elongation}
\end{eqnarray} 
It is remarkable that the elongation rates are almost the same for proton and
iron induced showers and nearly independent of the interaction
models. They are $58.5\pm 1.3$g/cm$^2$ per decade of energy for proton showers
and $60.9\pm 1.1$g/cm$^2$ per decade of energy for iron showers according to
the QGSJET model.  Any other values for E.R. observed from the data will
indicate a change in composition.

An important issue associated with measuring the elongation rate is
possible existence of a bias caused by tight cuts. To test this, we
compare the average $X_{max}$ as a function of energy using the simulated
events.  One set of data is based on all the CORSIKA sampled events and the
other is based only on those that trigger the detector, are able to be
reconstructed and pass the tight cut criteria. As shown
in FIG.\,\ref{ER}, no matter what
interaction model is used, we do not see any bias using
 the cuts set in the Table.\,\ref{cut_tab}. The thin
lines in the figure show the sampled events
and the circles and the squares show the
reconstructed results (see the figure legends for details).

\subsection{ The Change of Cosmic Ray Composition} 
\begin{figure}[t]
\epsfig{file=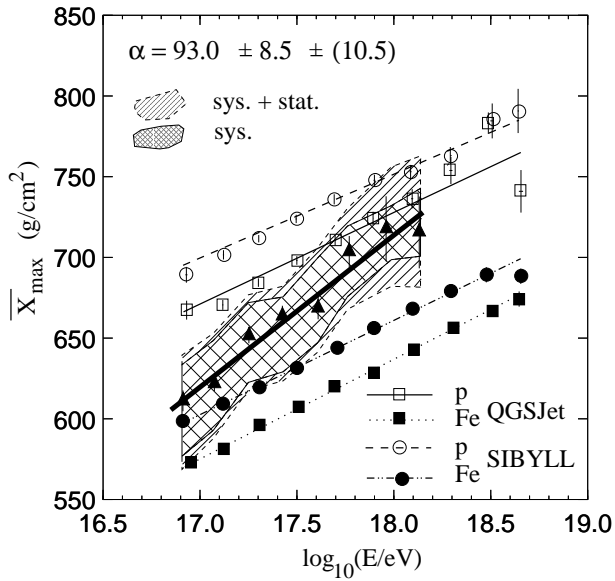, width=8cm} 
\caption{Average $X_{max}$ increasing with
energy.  Shaded areas and the thick line within the area represent HiRes data
and the best fit of the data respectively.  The closed triangles represent the
data set corresponding to the central values of the parameters in the
reconstruction.  The circles, squares and lines refer to the simulation
results.  See text for details. } 
\label{ER} 
\end{figure}

The HiRes/MIA experimental data
as shown in the FIG.\,\ref{ER} demonstrates an unambiguous change in average
$X_{max}$ with energy. This indicates a change towards a lighter mix of nuclei
in the average composition from $10^{17}$ to $10^{18}$eV. This indication of
change in composition can be evaluated by using the elongate rate measured in
the experiment, i.e.  
\begin{eqnarray} 
\alpha = 93.0 \pm 8.5 \pm (10.5)\hspace{0.5in} (\mbox{g/cm}^2), 
\label{ERnum} 
\end{eqnarray} 
where the number
in parentheses represents the systematic error discussed below. In
comparison with the predicted number for a unchanging or 
pure composition mentioned above, the
difference is larger than any known uncertainties. The
uncertainty in predicted elongation rate due to hadronic model 
dependence is small.  

We can estimate the change of the composition  
in the form of the  average logarithm of atomic 
number of the primary nuclei, $\Delta\overline{lnA}$, 
as $-1.5\pm 0.6$ over the energy range covered by the
data. This number is quite model independent assuming
 equal elongation rate for all different pure compositions. 
The systematic uncertainty in $\alpha$ is
included here. The absolute value of 
$\overline{lnA}$ is strongly model dependent as implied by FIG.\,\ref{ER}. 

\subsection{Uncertainties in $X_{max}$ and Energy}

Uncertainties in $X_{max}$ come both from choice of theoretical 
models and detector resolution. We first look into the uncertainty 
in the predictions. In FIG.\,\ref{ER}, the average difference in $X_{max}$ 
at any given energy between the predictions based on QGSJET and 
SIBYLL models is about 
25g/cm$^2$. Both are compatible with the data and both lead to the 
same qualitative conclusion of a lightening in the composition. 
However, the  value of 
 the average $lnA$ at any given energy is model dependent.

From the experimental point view, we have made a detailed effort to understand 
the systematic error in shower $X_{max}$ and energy.  
For $X_{max}$, we have considered systematic errors in the
atmospheric transmission of light and in the production of
Cerenkov light.  These are related since   
scattered Cerenkov light can masquerade as fluorescence light if
not accounted for properly.  For atmospheric scattering, there
is uncertainty in the aerosol concentration and its vertical distribution. 
The uncertainty, 
equivalent to one standard deviation 
with respect to the mean, is expressed as a range of possible
horizontal extinction lengths for aerosol scattering at 350\,nm
(taken as 11\,km to 17\,km based on measurements using Xenon flashers,  
\cite{flasher})
and a range of scale heights for the vertical
distribution of aerosol density above the mixing layer (taken as
0.6\,km to 1.8\,km).  For Cerenkov light production, we have
varied the angular scale for the Cerenkov emission angle over 
a one standard deviation equivalent.  
At ground level, we take the distribution as
an exponential function of the angle from the shower axis, with a
scale of $4.0\pm0.3^\circ$\cite{BLANCA}. Those uncertainties are shown by the 
shaded area in FIG.\,\ref{ER}. 

The systematic error in the energy is about 25\% and comes from fluorescence
efficiency uncertainty\cite{hires}, detector calibration
uncertainty\cite{calib} and the atmospheric corrections\cite{apj}. The first
two are intrinsically independent of the primary particle energy over this
range.  The fluorescence efficiency has been measured with an error of 10\%.
The percentage atmospheric corrections are also independent of energy because
the sample of showers is restricted to core locations within 2 km of the MIA
detector center. Therefore there is no significant atmospheric 
path length difference between an EAS and the 
detector for different energies. An energy independent
systematic fractional error in energy has no effect on the measured elongation
rate.  The magnitude of the systematic error in energy due to atmospheric
attenuation can be estimated by varying the atmospheric parameters over the
range described above. It is not greater than 10\%. The detector calibration
systematics is less than 5\%.

\subsection{$X_{max}$ Distribution } 
\begin{figure}[t]
\epsfig{file=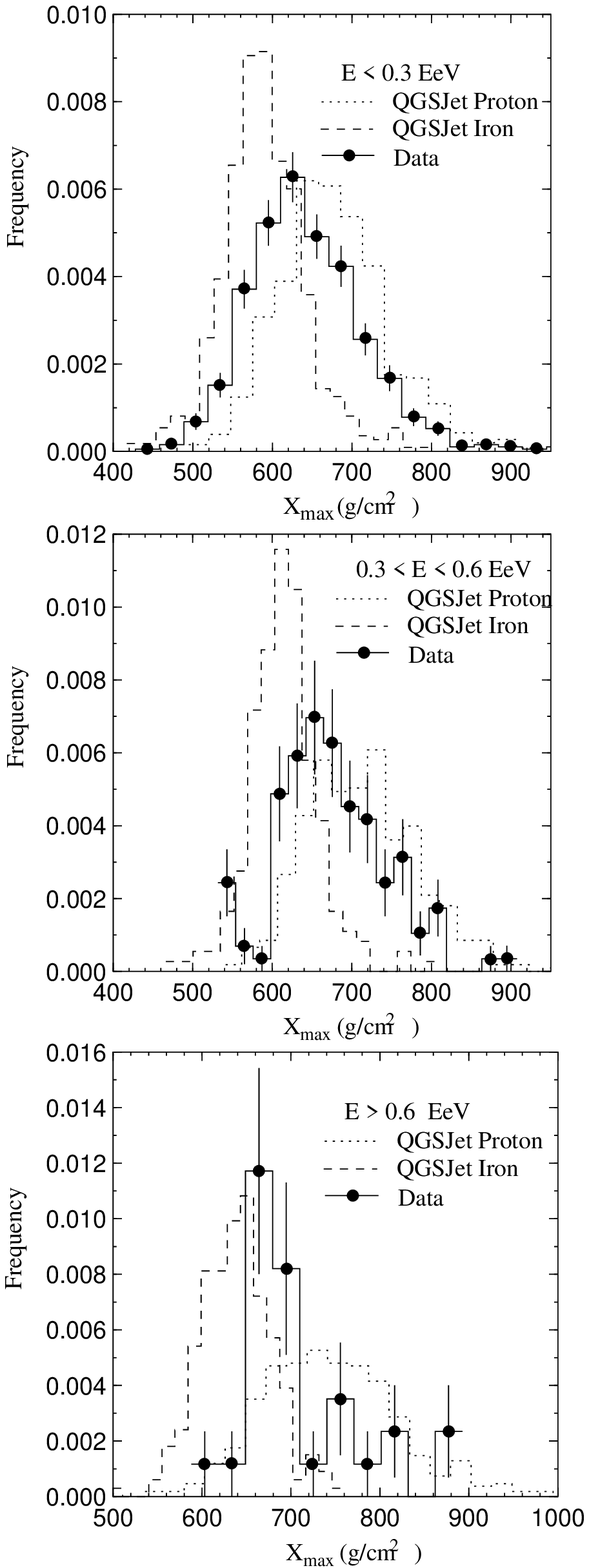, width=8cm}
\caption{$X_{max}$ distributions. Data from this experiment and simulated 
are compared. All distributions are normalized. }
\label{xm_dis}
\end{figure}

In FIG.\,\ref{xm_mc_dis}, we can see that the
fluctuations about the average $X_{max}$ for simulated proton showers is
larger than that for iron showers.  The fluctuations for both proton and iron
induced showers are too large to allow us to distinguish one from the other on
an event by event basis. 
However, we can determine the gross properties of the composition 
statistically.
In FIG.\,\ref{xm_dis} we plot the predicted distributions of
$X_{max}$ for proton and iron showers together with dotted and dashed lines,
respectively. The detector response has been folded into those
distributions. By studying those distributions in different energy ranges as
shown in the figure, one can compare the data to pure proton and pure 
iron composition distributions.
The data clearly requires a mixed composition of light and heavy particles
to account for the width and peak value of the $X_{max}$ distribution.

\subsection{Comparison with Previous Experiments}
\begin{figure}
\epsfig{file=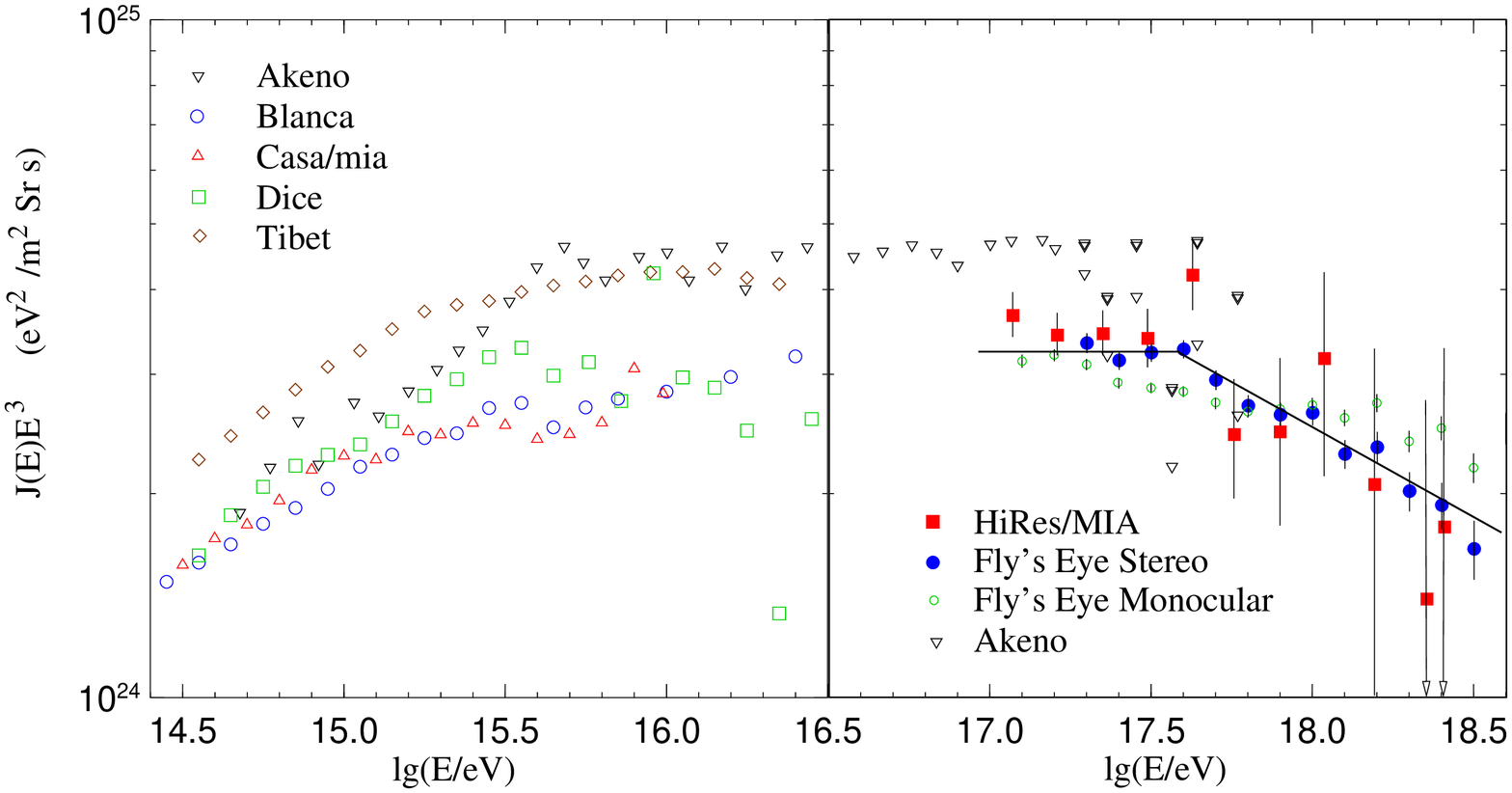, width=16cm}
\caption{Energy spectrum of cosmic rays from $10^{14.5}$ to $10^{18}$ eV.
$E^3$ is multiplied to the intensity.
Data in the vicinity of $3\times 10^{15}$ eV are adopted from 
\cite{BLANCA} (Blanca paper).}
\label{spe_all}
\end{figure}

The cosmic ray energy spectrum
 measured by all modern experiments are summarized 
in the FIG.\ref{spe_all} covering the whole energy range from 
$3\times 10^{14}$ to $3\times 10^{18}$ eV. The consistency between 
this experiment and the Fly's Eye stereo data has been discussed previously. 
We have shown the marginal confirmation of the break in the  
spectrum at $3\times 10^{17}$ eV there. By comparing with the observations
\cite{spectrum_all} in the ``knee'' region, we see that both the intensity
 and spectrum index imply a good continuity from the results at energy
lower than $3\times 10^{16}$ eV. We especially note that the change
in cosmic ray intensity around $3\times 10^{17}$ eV is comparable 
in power law index with 
the change that occurs around the ``knee''. A confirmation of this break
with  better statistics and similar energy resolution is important. 
All the other experimental
results are consistent with the  Akeno result: the spectrum follows a 
single index power law between $10^{16}$ and $10^{17}$ eV. 

The only existing experimental result based on direct measurements of shower
longitudinal development is that from Fly's Eye experiment. As a successor of
that experiment, the HiRes/MIA experimental result qualitatively 
supports old Fly's
Eye's result, i.e. there exists a trend in the composition of cosmic rays
to a lighter mix with energy. Quantitatively, they are
consistent with each other, taking into account the systematic errors in the
original Fly's Eye result of about 25 gm/cm$^2$ 
on individual $X_{max}$ measurements.
The elongation rate measured in HiRes/MIA experiment is marginally larger than
that in Fly's Eye\cite{apj} which is $78.9\pm 3.0$\,g/cm$^2$/decade where 
the quoted error is statistical only. We should also note that
experimentally measured 
elongation rates are not corrected for acceptance. Differences 
in acceptance for two experiments could introduce differences 
in elongation rates. The safest method is to compare each experiment to 
its own simulated proton and iron data sets.
The conclusion on the  composition of cosmic rays based on this kind of 
comparison is meaningful because the detector effects are counted in exactly
 the same way for both real and simulated events.  
For the present experiment, based on our simulation, we believe 
 that the detector biases for the elongation rate are minimal.
In summary, 
when all the errors are taken into account, the
results on $X_{max}$ distribution 
and elongation rates, from the two experiments are consistent, 
in spite of the differences in the analysis.

The other existing result on elongation rate in the same energy 
range is from the Yakutsk experiment\cite{Yukutsk}. The systematic 
error is not provided and the
method using a ground array experiment is more indirect than the present
experiment. Nevertheless, the results are marginally in agreement with the
result reported here.

\begin{figure} 
\epsfig{file=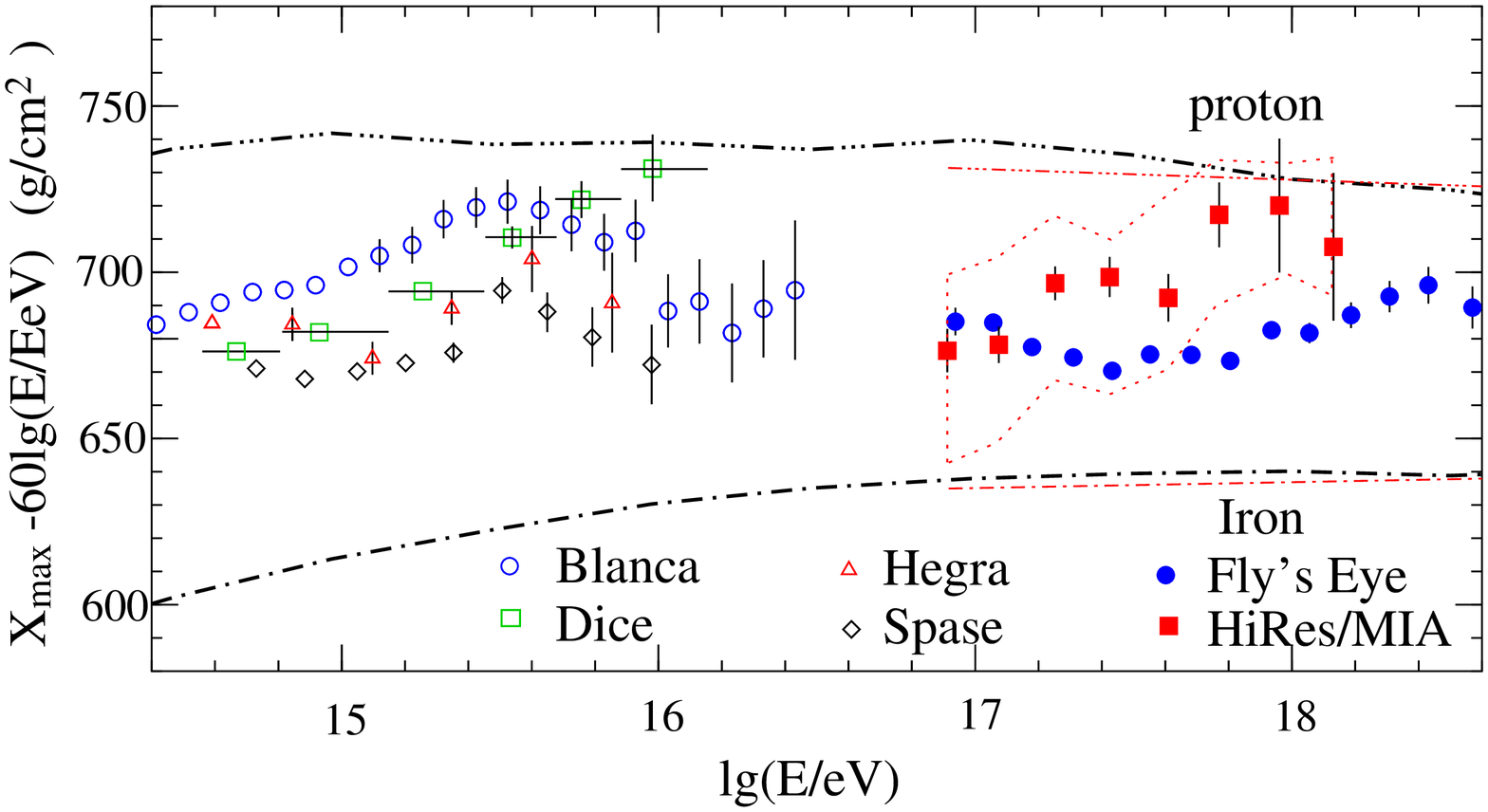, width=16cm} 
\caption{Average depth of shower maximum as a function of primary energy
of cosmic rays. An arbitrary elongation rate of 60 g/cm$^2$ is subtracted.
Again, the data in the lower energy region is adopted from 
\cite{BLANCA} (Blanca paper). Dashed lines with single-dots 
represent the simulated result from proton showers and dashed lines with
three-dots for the iron showers.  The thick lines are from \cite{BLANCA}
 and the thin lines from this experiment.}
\label{ER_all} 
\end{figure}

There are several measurements of elongation rate at lower energies, between
$10^{14}$ and few $10^{16}$ eV. These results and ours are shown in
FIG.\,\ref{ER_all} together. The trend of a changing cosmic 
ray composition shows a
pattern correlated with breaks in the energy spectrum.  It can be
characterized qualitatively as follows. There is a rather clear break around
$3\times 10^{15}$ eV which is related to the ``knee'' in the energy spectrum.
This break seems to be confirmed by several experiments \cite{BLANCA}. 
The elongation rate shows an increasingly heavy composition around this knee.  
Above
$3\times 10^{17}$ eV, the composition changes to a lighter mix. 
 It seems to be correlated to the spectral break observed 
by Fly's Eye experiment
which is marginally confirmed by this experiment.  Those experiments imply a
relatively unchanging region between $10^{16}$ and $10^{17}$ eV but no 
measurements of elongation rate exist between $10^{16.4}$ and $10^{17}$ eV.

\section{ Conclusion}
The HiRes/MIA hybrid experiment has measured  the cosmic
ray energy spectrum between $10^{17}$ and $3\times 10^{18}$. The spectral
index and intensity
 are given in Table.\,\ref{spectab}. The result is in agreement 
with the Fly's Eye experiment. This result 
marginally supports the Fly's Eye stereo observation of a break in the  
energy spectrum at $4\times 10^{17}$ eV.

The HiRes/MIA hybrid experiment confirms the Fly's Eye experimental result
that the elongation rate is different from simulation with an unchanging
composition. Modern hadronic interaction models and improved 
detector resolution in energy and $X_{max}$ do not change the original 
conclusion. Within errors,
the elongation rate observed in this experiment, $93.0\pm 8.5\pm
(10.5)$\,g/cm$^2$/decade, is consistent with previous experiments, such as
Fly's Eye and Yakutsk\cite{Yukutsk}.  While the conclusion regarding the
absolute value of $\overline{lnA}$ of the primary composition 
depends on the interaction
model used, this study shows that the elongation rate is stable
with respect to choice of models. In the light of this, 
the amount of the change in the average
composition, i.e. $\Delta\overline{lnA} = -1.5\pm 0.6$, is 
largely model independent, no matter what value
of $\overline{lnA}$ the change starts from.

Putting all experimental results together from $3\times 10^{14}$ 
to $3\times 10^{18}$ eV, we note that there seems to be a correlated
patterns in the energy spectrum and elongation rate, $X_{max}$ 
vs. energy, plot. Both measurements in energy spectrum and $X_{max}$ imply a 
continuity from lower to higher energies, with a flat bridge 
between $10^{16}$  and $10^{17}$ eV.

We note that following the break, the Fly's Eye experiment \cite{FEcomp}
reports a hardening of the spectrum near $5\times10^{18}$ eV. This has been
interpreted as evidence for the emergence of an extragalactic component above
a softer galactic component \cite{FEcomp}.  A change from a heavy to a light
composition in this energy region also gives support to a changing origin for
those cosmic rays.  The lack of a strong galactic anisotropy at the highest
energies would also rule out galactic sources for energetic 
protons \cite{Alfred}.
A number of new experiments, such as HiRes, the Pierre Auger Project, the
Telescope Array, EUSO and OWL will address this issue.

\section{Acknowledgements}
We acknowledge the assistance of the command and staff of Dugway Proving 
Ground. This work is supported by the National 
Science Foundation under contract No. PHY-93-21949, PHY-93-22298 and 
U.S. Department of Energy and the Australia Research Council.

\end{document}